# Lattice Misfit Measurement in Inconel 625 by X-Ray Diffraction Technique

P.Mukherjee, A.Sarkar and P.Barat.
**Variable Energy Cyclotron Centre, 1/AF Bidhannagar,
Kolkata – 700 064, India.**

T.Jayakumar and S. Mahadevan.
**Indira Gandhi Centre for Atomic Research, Kalpakkam – 603 102, India.**

Sanjay K. Rai
**Centre for Advanced Technology, Indore – 452013, India.**

## Abstract

Determination of lattice misfit and microstructural parameters of the coherent precipitates in Ni based alloy Inconel-625 is a challenging problem as their peaks are completely overlapping among themselves and also with the matrix. We have used a novel X-ray diffraction technique on the bulk samples of Inconel 625 at different heat-treated conditions to determine the lattice parameters, the lattice misfit of the coherent precipitates with the matrix and their microstructural parameters like size and strain.

## Introduction

Nickel-based superalloy Inconel 625 is widely used in aeronautic, aerospace, marine, chemical and petrochemical industries [1]. It is also a candidate material used for reactor-core and control rod components [2] in pressurized water reactors and in ammonia cracker units for heavy water production and other petrochemical industries. The alloy possesses high strength and toughness over a wide range of temperature and is endowed with excellent fatigue strength, oxidation resistance and corrosion resistance [3]. This alloy was initially designed keeping in mind to provide high strength due to solid solution strengthening of nickel-chromium matrix by molybdenum and niobium. However, later it was found that a large number of ordered intermetallic phases precipiate in the alloy during service or on aging treatment at high temperature in the range of 823K to 1023K. The previous studies on this alloy [2,4] revealed that depending upon the time and temperature of exposure $M_{23}C_6$, $M_6C$, MC carbide ('M' is rich in Ni, Nb and Mo), $\gamma''$ [$Ni_3$(Nb, Al, Ti)] and $Ni_2$(Cr, Mo) precipitate in this alloy system. Among these precipitates, $\gamma''$ has body centered tetragonal $DO_{22}$ structure and $Ni_2$(Cr, Mo) possesses orthorhombic $Pt_2Mo$ type structure.

Since the precipitates of $\gamma''$ and $Ni_2$(Cr, Mo) are generally small in size (~10 to 20 nm) and coherent with the matrix, a major source of strengthening in this alloy is the coherency strain. The lattice mismatch between the precipitates and the matrix and between the two types of precipitates plays an important role in the process of co-precipitation. The coherency strain also influences the coarsening behaviour of the precipitates and the long term behaviour of the microstructure and mechanical properties at elevated temperature [5].

Lattice misfit measurement between the precipitates and the matrix is common in Nickel base superalloys [6-8]. Various techniques like X-ray diffraction (XRD), neutron diffraction (ND) and Convergent Beam Electron Diffraction (CBED) have been used for this purpose. Measurements were performed on the samples where the precipitates are either embedded in the matrix or after extracting the precipitates by dissolving the matrix [9]. Further, the presence of more than one type of precipitates with peaks overlapping with the peaks of the $\gamma$ matrix requires a high resolution measurement technique if bulk specimens are used. Transmission electron microscopy (TEM) is the most suitable and direct technique to characterise the precipitates in nickel base superalloys [1,2]. However, only a miniscule volume of the material is examined under TEM within a thin film (~100 $\mu m^3$) and hence it may not depict the microstructure of the bulk. On the contrary, XRD technique [10] gives the microstructural information in a statistical manner, averaged over a volume of $10^9$ $\mu m^3$. Moreover, the analysis is much easier, reliable and quick and the specimen preparation requires very little time. The prime objective of this study is to establish that XRD technique can be reliably used to characterise the crystal structure and the microstructural parameters of fine coherent precipitates.

Here, we report the studies carried out on the bulk samples of Inconel 625 at different heat-treated conditions using XRD technique, for determination of the lattice parameters of the precipitates [$\gamma''$ and $Ni_2(Cr, Mo)$], their size and the lattice misfit with the matrix due to the presence of these coherent precipitates. The separation of the overlapping peaks (hidden peaks) was done using a mathematical method [11] available in peak separation software. From the extent of broadening of the each separated peak, the size of the precipitates was evaluated [12].

**Experimental**

The current investigation has been conducted using Inconel 625 retrieved from an ammonia cracker plant after 60000h of service at approximately 873K. The blanks of 9mmx10mmx60mm were cut from the service exposed (SE) tubes and they were then thermally aged at temperatures 923K for one hour (SE 923K) and 1123K for one hour (SE 1123K).

The chemical composition (in wt %) of the alloy investigated is as follows: Cr-21.7, Fe-3.9, Mo-8.8, Nb-3.9, C-0.05, Mn-0.14, Si-0.15, Al-0.17, Ti-0.23, Co-0.08 and rest Ni. Specimens were taken for X-ray diffraction analysis from the service exposed (SE) as well as from the heat-treated tubes.

XRD profiles of the samples were recorded using MAC Science MXP18 X-ray diffractometer using $CuK_\alpha$ radiation. The scan was performed in the angular range of 25 to 100 degrees with a step size of 0.02 degrees. The time per step was 4 seconds. All the specimen were polished and electrolytically etched before carrying out the XRD scan.

## Method of Analysis

Diffraction of radiation from matter corresponds to a Fourier transform from real space to momentum space; hence the XRD pattern of a sample represents a complete mapping of its crystal structure and microstructure in the momentum space. In most investigations, structural (crystal) information is extracted from the diffraction pattern namely the angular positions and intensities of the Bragg peaks. However, in the present study, we are interested in the microstructural features. Generally, the broadening of the diffraction peak is due to the instrumental broadening, broadening due to the particle size and microstrain arising due to the presence of small and coherent particles. The contributions from each of these effects are convoluted causing an overall broadening of the diffraction peaks.

The microstructure of SE Inconel 625 consists of $\gamma$, $\gamma''$ and $Ni_2(Cr, Mo)$ [1,2,13]. Fig.1a shows XRD profile of SE Inconel 625. The single peak which is observed in the $2\theta$ range 41.5° – 45.3° essentially consists of three overlapping peaks i.e. (111) peak of $\gamma$, (111) peak of $Ni_2(Cr,Mo)$ and (031) peak of $\gamma''$. Similarly, the overlapping peaks in the other 2θ range of the XRD profile of SE Inconel 625 are also shown in Fig. 1a. To determine the lattice parameters accurately for each of these three phases and also to calculate the misfit values among the phases, proper separation of the overlapping peaks is essential. Several mathematical methods exist to find the hidden peaks. Some of these methods such as (1) the residual method, (2) the second derivative method and (3) the deconvolution method, are available in a peak separation software namely, PeakFit version 4.0 for windows [11]. We have used the first method in our study for analysis of the diffraction peaks of the specimens of Inconel 625. In this method, a residual is simply the difference in the y values between a data point and the sum of component peaks evaluated at the data point's x value. By placing peaks in such a way that the total area of the individual peaks equals to the area of the experimentally observed peak, and hence the hidden peaks are revealed by residual method.

Before, extracting the information about the microstructure from the XRD profile, it is necessary to correct the instrumental effect. The actual diffraction profile, which is basically a convolution of the instrumental broadening and total sample broadening (i.e. broadening due to size and strain), can be represented as:

$h(x) = g(x) * f(x)$

where, $h(x)$, $g(x)$ and $f(x)$ represent the actual diffraction profile, the instrumental broadening and sample broadening respectively.

A careful scan of a suitable standard sample showing minimal physical broadening define the instrumental broadening $g(x)$. For this purpose, we have taken a silicon sample as standard which is almost free from defects. The diffraction profile from the silicon sample is fitted with pseudo-Voigt (pV) function [14]. The Instrumental Resolution Function (IRF) is then determined using the Caglioti relation [15],

$(FWHM)^2 = U tan^2\theta + V tan\theta + W$

and  $1-\eta = a + b(2\theta)$

where *FWHM* is the Full Width at Half Maxima, *U, V, W*, *a, b* are the instrument parameters and $\eta$ is the mixing parameter of the Cauchy and the Gaussian part.

The instrumental resolution function can be expressed by:

$PV(x) = I_0[(1-\eta)exp[-ln(2)x^2/\omega^2] + \eta 1/1 + x^2/\omega^2]$

Where $I_0$ is the peak intensity, $x = 2\theta - 2\theta_0$ and $\omega$ = *HWHM i.e.* Half Width at Half Maximum.

Using this function, the peaks are generated due to instrumental contributions at the Bragg peak positions and the corresponding values of FWHM are then incorporated in the program PeakFit 4.0 to deconvolute the instrumental effect.

A Voigt function was used for approximating the peak profile. Both the width and the shape of the peak were allowed to vary during the actual peak fitting operation.

The size and the strain of the γ″ and $Ni_2(CrMo)$ precipitates cause broadening of the diffraction peaks. From the broadening of the individual hidden peak for each phase, the particle size and microstrain values were calculated using the relation [11],

$$\frac{\beta \cos\theta}{\lambda} = \frac{1}{D_v} + 2\varepsilon\left(\frac{2\sin\theta}{\lambda}\right)$$

The plot [Williamson – Hall plot (12)] of $\left(\frac{\beta \cos\theta}{\lambda}\right)$ versus $S = \left(\frac{2\sin\theta}{\lambda}\right)$ gives the value of microstrain from the slope and domain size from the ordinate intersection.

**Results and discussion**

A typical example of the fitted peaks for the sample SE 923K is shown in Fig 2. Three overlapping peaks namely (111) γ, (111) peak of $Ni_2(CrMo)$ and (031) peak of γ″ at 2θ values of 43.375, 43.530 and 43.175 respectively, could be fitted to the peak which lies between the 2θ range 41.5°-45.3°, using the program PeakFit version 4.0. Similarly, the other peaks i.e. (200) of γ overlaps with (002) peak of $Ni_2(Cr,Mo)$ phase, (220) peak of γ overlaps with (132) peak of $Ni_2(Cr,Mo)$ phase and (060) of γ″ phase and (311) peak of γ overlaps with (033) of γ″ phase. All these peaks were fitted using the program to reveal the contributions of the individual phases to the intensity values of the experimentally observed peak after the correction of the instrumental broadening. The fitted curves show that they lie within the 99% confidence limit. The results of the individual peak fitting from the diffraction profiles of the SE specimen and the specimens which were heat treated at 923K for 1hour (SE 923K) and 1123K for 1 hour (SE 1123K), are compiled in Table 1, Table 2 and Table 3 respectively. The lattice parameters of each phase were then calculated from the hidden peaks using least square fitting considering the appropriate crystal structure and the results are shown in Table 4. The lattice misfit was calculated according to the equation [15],

$$\delta = \frac{a_1 - a_2}{\frac{1}{2}(a_1 + a_2)}$$

where δ is the lattice mismatch, $a_1$ is the lattice parameters of the precipitate phase and $a_2$ is the lattice parameters of the matrix phase. The results are included in Table – 4.

The values of the microstrain and the size of the particles of the phases $\gamma''$ and Ni$_2$(Cr,Mo) obtained from Williamson-Hall plot are shown in Table-5.

From, Table-1, it is clear that both $\gamma''$ and Ni$_2$(CrMo) precipitates are present in the matrix of $\gamma$ for the SE specimen. The values of FWHM, d-spacing and the intensity ratio estimated from the area under the individual peaks (I/I$_1$ from fit) are shown in Table-1. The results are qualitatively consistent with the reported TEM observation regarding the presence of both $\gamma''$ and Ni$_2$(Cr,Mo) phases [2].

The post service ageing treatment at 923K for 1 hour would have initiated the dissolution of Ni$_2$(Cr,Mo) and $\gamma''$. There is a trend in the reduction of the intensity ratio in most of the peaks of both the phases in the sample SE 923K (Table-2) as compared to the intensity ratio of these peaks in SE specimen (Table-1).

On the other hand, the fitted results of the diffraction peaks for the sample SE 1123K showed the existence of a single phase which is definitely $\gamma$ phase. This implies the complete dissolution of Ni$_2$(Cr,Mo) and $\gamma''$ at 1123K. This observation is in agreement with the reported literature [2]. It is to be noted that presence of carbides in the matrix could not be detected by XRD in all the samples.

Table 4 indicates the values of lattice parameters obtained from the analysis. The values of the lattice parameters of each phase match with the reported values [16, 17]. It is seen from Table-4, that the lattice misfit between $\gamma$ and $\gamma''$ was found to be positive for SE and SE 923 K as there is a lattice expansion in both the cases. The misfit values of $\gamma$ and $\gamma''$ were found to be 0.224 for SE and 0.290 for SE 923 K. The lattice parameter of $\gamma$ phase of the SE sample was found to be 3.6073Å and the partial dissolution of $Ni_2(Cr,Mo)$ and $\gamma''$ changes the value to 3.6092 Å. This may be attributed to the dissolution of Cr and Mo, back into the matrix. Complete dissolution of these phases after post service ageing treatment at 1123K for 1 hour further increases the lattice parameter. The lattice misfit of $\gamma$ phase and $Ni_2(Cr,Mo)$ were also found to be positive and the values were 0.063 and 0.042 for SE and SE 923K samples.

Table 5 gives the values of the particle size of the phases $Ni_2(Cr,Mo)$ and $\gamma''$ and the corresponding strain resulted due to the presence of these coherent precipitates. In SE specimen, presence of $\gamma''$ created coherency strain in the matrix of $\gamma$ due to the lattice mismatch. The values of strain decrease after post service ageing treatment which caused the dissolution of the phases at high temperature.

## Conclusion

X-ray diffraction data obtained from Inconel 625 at different heat-treatment conditions could be used reliably to determine the lattice parameters of $Ni_2(Cr,Mo)$ and $\gamma''$ in the matrix of $\gamma$. The mismatch between the lattice parameters between $\gamma''$-$\gamma$ and $\gamma$-$Ni_2(CrMo)$ could be calculated by separation of the hidden peaks using Residual method. Microstructural parameters like precipitate size and microstrain within the matrix resulting from the precipitates $Ni_2(CrMo)$ and $\gamma''$ could be well characterised from the peak broadening of these individual peaks. Thus, X-ray diffraction technique can be a very good tool to study the microstructure of the coherent precipitates.

## Acknowledgements

Authors are grateful to Prof. Bikash Sinha, Director, Variable Energy Cyclotron Centre and Dr. D.K.Srivastava, Group Director, Physics Group, Variable Energy Cyclotron Centre, Dr. Baldev Raj, Director, Indira Gandhi Centre for Atomic Research, Kalpakkam, Dr. S.L.Mannan, Group Director, Metallurgy and Materials Group and Mr P. Kalyansundaram, Associate Director, Inspection Technology Group, IGCAR for constant encouragement and support. Authors are thankful to Dr. Anish Kumar of Non Destructive Evaluation Division, IGCAR, for many useful discussions.

**Table-1: Values obtained from peak fitting of X-ray diffraction data for overlapping peaks of different phases for SE samples**

| Index of plane (h k l) | 2θ (°) | FWHM (°) | d- spacing (Å) | I/I$_1$ (from fit) (%) |
|---|---|---|---|---|
| γ-matrix | | | | |
| 1 1 1 | 43.380 | 0.252 | 2.086(±0.002) | 75.63 |
| 2 0 0 | 50.492 | 0.446 | 1.807(±0.004) | 87.01 |
| 2 2 0 | 74.356 | 0.485 | 1.276(±0.003) | 83.64 |
| 3 1 1 | 90.287 | 0.786 | 1.087(±0.005) | 90.56 |
| Ni$_2$(Cr,Mo) | | | | |
| 1 1 1 | 43.741 | 0.287 | 2.069(±0.003) | 8.24 |
| 0 0 2 | 50.554 | 0.491 | 1.805(±0.004) | 12.99 |
| 1 3 2 | 74.584 | 0.350 | 1.272 (±0.005) | 7.50 |
| γ″ | | | | |
| 0 3 1 | 43.202 | 0.332 | 2.094(±0.002) | 16.13 |
| 0 6 0 | 74.185 | 0.450 | 1.278 (±0.004) | 8.85 |
| 0 3 3 | 89.963 | 0.936 | 1.090(±0.005) | 9.44 |

**Table-2: Values obtained from peak fitting of X-ray diffraction data for overlapping peaks of different phases for SE 923K samples**

| Index of plane (h k l) | 2θ (°) | FWHM (°) | d- spacing (Å) | $I/I_1$ (from fit) (%) |
|---|---|---|---|---|
| γ-matrix | | | | |
| 1 1 1 | 43.375 | 0.123 | 2.085(±0.002) | 78.28 |
| 2 0 0 | 50.489 | 0.364 | 1.807(±0.003) | 89.70 |
| 2 2 0 | 74.349 | 0.313 | 1.276(±0.004) | 84.15 |
| 3 1 1 | 90.212 | 0.427 | 1.088(±0.005) | 94.48 |
| $Ni_2(Cr,Mo)$ | | | | |
| 1 1 1 | 43.530 | 0.134 | 2.078(±0.002) | 8.69 |
| 0 0 2 | 50.568 | 0.772 | 1.805(±0.004) | 10.30 |
| 1 3 2 | 74.498 | 0.384 | 1.274(±0.005) | 6.88 |
| $\gamma''$ | | | | |
| 0 3 1 | 43.175 | 0.513 | 2.095(±0.003) | 13.01 |
| 0 6 0 | 74.136 | 0.336 | 1.278(±0.004) | 8.96 |
| 0 3 3 | 89.887 | 0.279 | 1.091(±0.005) | 5.52 |

**Table-3: Values obtained from peak fitting of X-ray diffraction data for overlapping peaks of different phases for SE 1123K samples**

| Index of plane ($h\ k\ l$) | $2\theta$ (°) | FWHM (°) | $d$- spacing (Å) | $I/I_1$ (from fit) (%) |
|---|---|---|---|---|
| $\gamma$-matrix | | | | |
| 1 1 1 | 43.335 | 0.101 | 2.087($\pm$0.002) | 100 |
| 0 0 2 | 50.551 | 0.190 | 1.805($\pm$0.003) | 100 |
| 2 2 0 | 74.282 | 0.137 | 1.276($\pm$0.003) | 100 |
| 3 1 1 | 90.164 | 0.187 | 1.088($\pm$0.005) | 100 |

**Table-4: Values of lattice parameters and lattice misfit of different phases**

| Sample | Lattice parameter (Å) | | | Lattice misfit (%) | |
|---|---|---|---|---|---|
| | γ Phase | $Ni_2(Cr,Mo)$ | $γ''$ Phase | $γ/γ''$ | $γ/Ni_2(Cr,Mo)$ |
| Service Exposed | a=3.6073 (±0.0005) | a=2.6916(±0.0002) b=7.3168(±0.0008) c=3.6096(±0.0003) | a=3.6154(±0.0008) c=7.6699(±0.0009) | (+)0.224 | (+)0.063 |
| SE 923 K 1h | a=3.6092 (±0.0004) | a=2.7156(±0.0003) b=7.1981(±0.0007) c=3.6107(±0.0004) | a=3.6197(±0.0007) c=7.6746(±0.0009) | (+)0.290 | (+)0.042 |
| SE 1123 K 1h | a=3.6108 (±0.0006) | | | | |

**Table 5: Size of the precipitates and microstrain values**

| Sample/Phases | Size(nm) | Strain($10^{-3}$) |
|---|---|---|
| **Service Exposed** | | |
| $\gamma$ | --- | 3.7($\pm$0.6) |
| $Ni_2(Cr,Mo)$ | 21.0($\pm$2.0) | 0.5 ($\pm$0.06) |
| $\gamma''$ | 35.0 ($\pm$4.0) | 2.3($\pm$0.1) |
| | | |
| SE 923 K 1h | | |
| $\gamma$ | --- | 2.8($\pm$0.2) |
| $Ni_2(Cr,Mo)$ | 9.6($\pm$1.0) | negligible |
| $\gamma''$ | 20.0($\pm$2.0) | negligible |
| | | |
| SE 1123 K 1h | | |
| $\gamma$ | --- | 0.9($\pm$0.1) |

# Figure Captions

1. X-ray Diffraction Patterns of (a) Service Exposed (SE), (b) SE 923K for 1 hour, (c) SE 1123K for 1 hour.

2. A typical example of the fitted peaks for the sample SE 923 K.

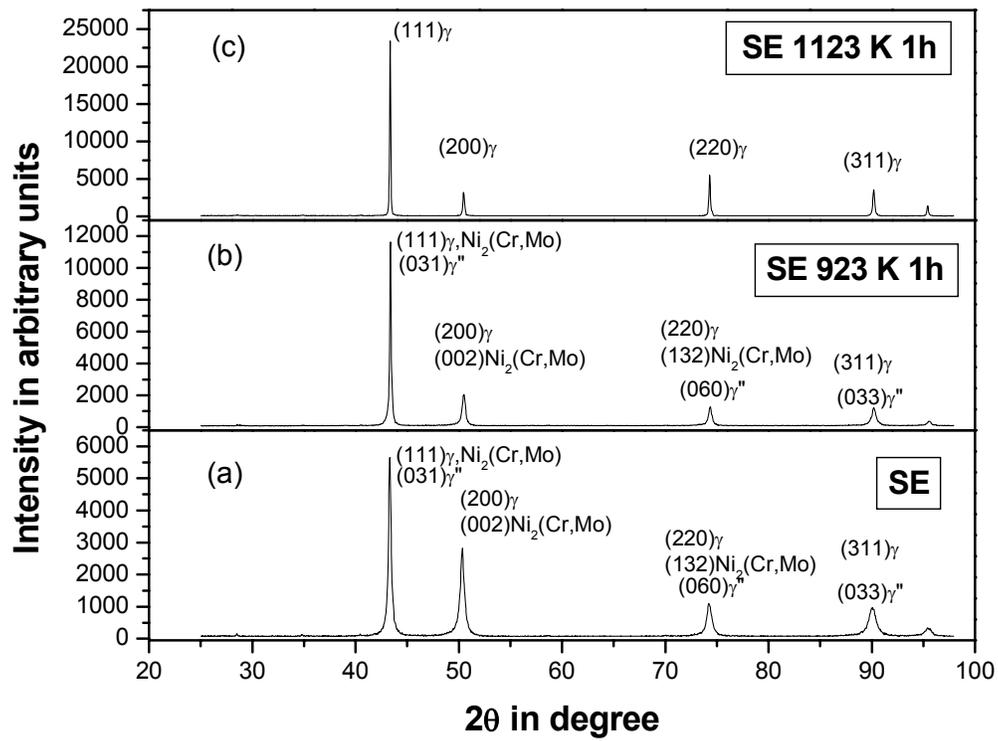

**Fig. 1**

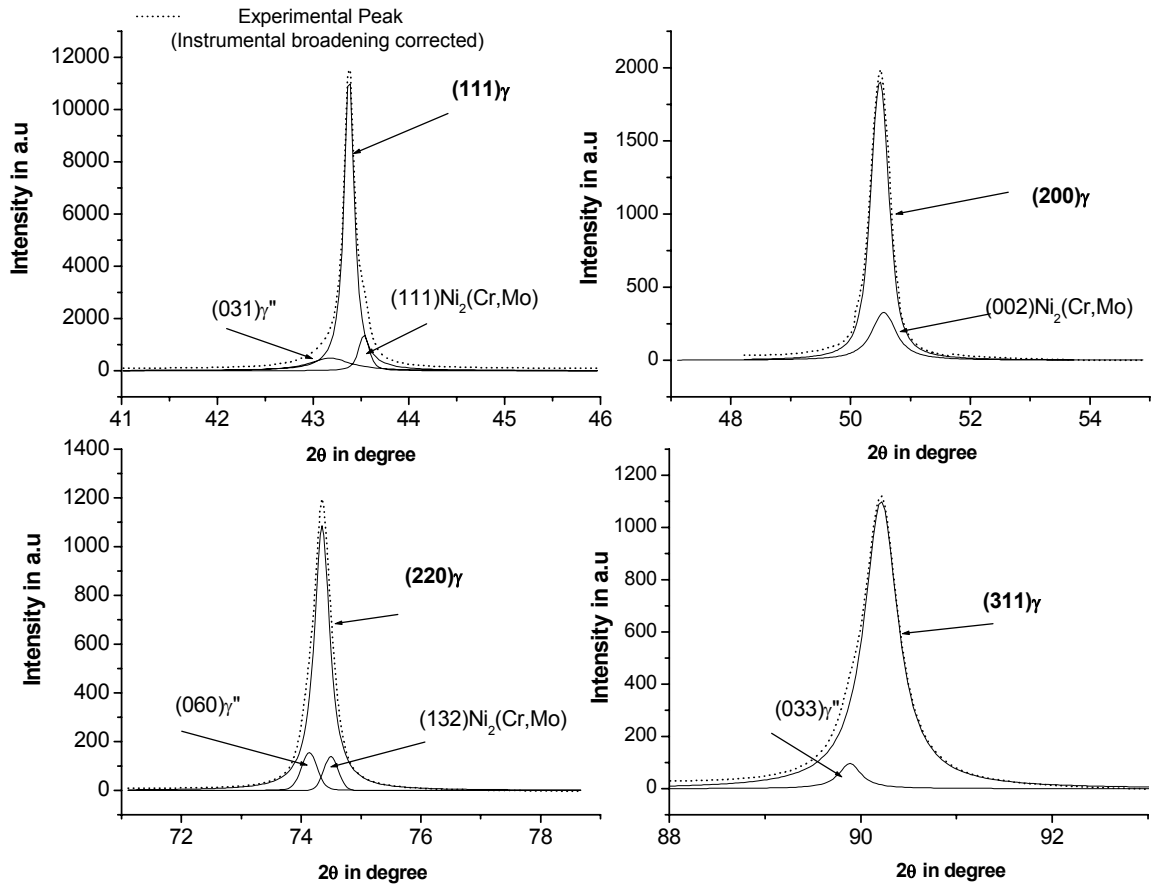

**Fig. 2**